\documentclass[10pt,twocolumn]{article} 
\usepackage{graphicx}
\usepackage{url,hyperref}
\usepackage{amsmath}
\usepackage{amssymb}
\usepackage{amsfonts}
\usepackage{algorithmic}
\usepackage{textcomp}
\usepackage{xcolor}
\usepackage{siunitx}
\usepackage{todonotes}
\usepackage{float}
\usepackage{caption}
\usepackage{subcaption}
\usepackage{soul}
\usepackage{array}  
\usepackage{booktabs} 
\usepackage{titlesec}

\font\tenhv  = phvb at 10pt

\def\@maketitle
   {
   \newpage
   \null
   \vskip .375in 
   \begin{center}
      {\Large \bf \@title \par} 
      \vspace*{24pt} 
      {
      \large 
      \lineskip .5em
      \begin{tabular}[t]{c}
         \@author 
      \end{tabular}
      \par
      } 
      \vskip .5em 
      {
       \large 
      \begin{tabular}[t]{c}
         \@affiliation 
      \end{tabular}
      \par 
      \ifx \@empty \@email
      \else
         \begin{tabular}{r@{~}l}
            E-mail: & {\tt \@email}
         \end{tabular}
         \par
      \fi
      }
      \vspace*{12pt} 
   \end{center}
   } 

\def\abstract
   {%
   \centerline{\large\bf Abstract}%
   \vspace*{12pt}%
   }

\def\affiliation#1{\gdef\@affiliation{#1}} \gdef\@affiliation{}

\def\email#1{\gdef\@email{#1}}
\gdef\@email{}

\long\def\@makecaption#1#2{
   \vskip 10pt
   \setbox\@tempboxa\hbox{\tenhv\noindent #1.~#2} 
   \setlength{\@ctmp}{\hsize}
   \addtolength{\@ctmp}{-\@figindent}\addtolength{\@ctmp}{-\@figindent} 
   \ifdim \wd\@tempboxa >\@ctmp
      \begin{list}{}{\leftmargin\@figindent \rightmargin\leftmargin} 
         \item[]\tenhv #1.~#2\par
      \end{list}
   \else
      \hbox to\hsize{\hfil\box\@tempboxa\hfil} 
   \fi}

\begin{document}

\title{Numerical Studies of Optimized Designs for Carbon Nanotube\\ Microstrip Antennas}

\author{HeeBong Yang$^{1}$, Aaron Gross$^{1}$, and Na Young Kim$^{1,2}$\\
\\
$^1$Institute for Quantum Computing, Waterloo Institute for Nanotechnology,\\ Department of Electrical and Computer Engineering, University of Waterloo,\\ Waterloo, ON N2L 3G1, Canada \\
$^2$Department of Physics and Astronomy, Department of Chemistry,\\ University of Waterloo,\\ Waterloo, ON N2L 3G1, Canada \\
\\
}
\date{}

\maketitle
\thispagestyle{empty}

\begin{abstract}
We perform systematic numerical simulations for carbon nanotube (CNT) film microstrip antennas to fabricate flexible and durable applications in terms of various device design parameters. The selection of appropriate materials for conductive films and a substrate of the conformable and robust microstrip antennas are crucial to increase the radiation efficiency and to reduce the losses while maintaining the mechanical properties. CNTs have been spotlighted as a promising nano-material, exhibiting excellent electrical and mechanical performances as desirable features for microwave wearable devices. Considering the material properties of the conductor and the substrate, we examine the possible ranges of the CNT film conductivities, conductive film thickness, and a dielectric constant and thickness of a substrate. Furthermore, we model non-uniform spatial distributions of conductivity in the CNT film to assess their impact on the antenna performance. Our extensive studies of material constants and conductivity spatial patterns propose design guidelines for optimal microstrip antennas made of CNT conductive films operating in microwave frequencies.
\end{abstract}

\section{Introduction}
Microstrip or patch antennas have been actively developed and widely deployed ever since G. A. Deschamps first demonstrated it in 1953 \cite{Deschamps_1953, Bernhard_2003}. Their versatility is attributed to several advantageous features: simple planar structures for scalability with antenna arrays, mature fabrication processes, cost-effectiveness, compact 
form factors, low weight, and  straightforward installation. In particular, the explosive needs of mobile phones and the boom of high-speed wireless communications continue to ask huge demands of innovative microstrip antennas based on nanomaterials.

A microstrip antenna consists of a dielectric substrate sandwiched by a ground plane at the bottom and a small (traditionally, rectangular-shaped) conductive patch on the top, which radiates. When we design microstrip antennas, we pay attention to two aspects: (1) right materials for substrates and conductors, and (2) geometric  dimensions. First, for the antenna substrate materials, a dielectric constant ($\epsilon_{\text{r}}$) and thickness ($h$) 
are studied. Several different materials have been studied as the antenna substrates: paper, textile, polyethylene terephthalate (PET), kapton polyimide, polydimethylsiloxane (PDMS), and nickel aluminate \cite{Anagnostou_2010,Locher_2006,Tighezza_2018,Khaleel_2012,Ramli_2017,Rahman_2019}. In particular,  polytetrafluorethlene (PTFE) and FR4 have been commonly chosen for RF and microwave antennas, and their $\epsilon_{\text{r}}$ are 2.0 - 2.3 and 3.9 - 4.7, respectively. The impact of $\epsilon_{\text{r}}$ values on the antenna performance is huge since it affects the radiation efficiency ($\eta$) in that the smaller $\epsilon_{\text{r}}$ generally increases $\eta$ \cite{Duffy_1996}. On the other hand, the higher $\epsilon_{\text{r}}$ is advantageous to make a thinner substrate \cite{Roy2013}.  Next, thickness of the substrate ($h$) is another essential design factor. It is generally known that thicker substrates enhance $\eta$ \cite{Pozar_1992}, but $h$ values are closely related to antenna operation frequency and material properties. For some materials like paper or fabric, $h$ is not necessarily thick. In recent times, there are growing demands for flexible antennas as an alternative to the conventional rigid substrates, especially in flexible, wearable, and stretchable communication devices and electronics \cite{Hassan_2021}. 

Parallel to the substrate material search, a variety of novel conductive materials, instead of the traditionally used copper (Cu), have been recently tested for the metallic patch itself such as graphene-based conductive films~\cite{Sajl_2015}, silver and carbon-based conductive inks~\cite{Mesquita_2019}, and carbon nanotubes (CNTs) \cite{Verma2013, Bengio_2019, Suryanarayana2021}. Note that all of these materials exhibit a wide range of conductivity values, which can directly influence both structure geometries and device figure of merits, for example, radiation efficiency $\eta$ or reflection loss. Numerical simulations were performed to assess the effect of materials with lower conductivity ($\sigma$) than Cu in wearable microstrip antennas~\cite{Yilmaz_2008}, and the thickness variation effect of the conductive films was numerically evaluated with polypyrrole conductive polymer \cite{Verma_2010}.  While the design of microstrip antennas with conventional conductors such as Cu is straightforward and well developed ~\cite{Balanis_2016}, we have not seen quantitative and analytical explanations to the role of the $\sigma$ values in the antenna design. It is often neglected with an ambiguous and qualitative comment of ``sufficient $\sigma$ is required.".

High conductivity is a basic and necessary trait of the conductive material for the flexible microstrip antenna. The durability is also an essential material property against strains or stresses from the action of stretch or compression. In these regards, CNT is an appropriate material candidate because the CNT possesses not only superb electrical attributes of high conductivity and high mobility \cite{Wang_2018} but also outstanding mechanical characteristics with high Young's modulus \cite{Forro_2002} and tensile strength \cite{Takakura2019}. Recently, E. Amram Bengio \textit{et al.} demonstrated microstrip antennas made of aligned single-walled carbon nanotube (SWCNT) films~\cite{Bengio_2019}. An individual SWCNT is indeed a good conductor with $\sigma_{\text{CNT}} \sim$ $10^6 - 10^7$ S/m comparable to Cu ($\sigma_{\text{Cu}} \sim 5.96 \times 10^7$ S/m) \cite{Wang_2018,Earp_2020}. However, we assume that SWCNT films for antenna devices are treated as macroscopic conducting materials because  the  wavelengths of the propagation waves are on the order of centi-meters (for 5 GHz) which is much longer  than the typical CNT lengths around several micron. Therefore we rather focus on the large scale behavior by  neglecting the microscopic modeling in our studies. In comparison to an individual CNT only, the CNT films often exhibit lower $\sigma$ owing to the mixture of different SWCNTs and their random orientation in films. The authors made both aligned and  non-aligned SWCNT thin films to make 5 GHz, 10 GHz, and 14 GHz antenna devices whose sheet resistance varies between 0.3 $\Omega.\text{sq}^{-1}$ and 1.3 $\Omega.\text{sq}^{-1}$, which are related to $\sigma$. In order to see the relations of $\sigma$ and antenna performances, the authors measured a voltage standing wave ratio (VSWR) and $\eta$ \cite{Bengio_2019}. With respect to Cu antennas, the CNT antennas exhibit similar $\eta$ of 80-90 \% over the sheet resistance value and thinner films less than the skin depth at 14 GHz. VSWRs for CNT devices are better than that of Cu devices by 20 \%. However, the aligned CNT devices do not perform exceptionally better than the non-aligned as-grown CNT film antennas. These results indicate that $\sigma$ is not the only key parameter to determine the radiation efficiency property of the microstrip antennas although higher $\sigma$ has a low ohmic loss, consequently leading to higher $\eta$.

To explain these experimental results, we also look into the conductor skin depth ($\delta$) carefully with regards to the conductor film thickness $t$, and the quality factor. The skin depth effect is significant in AC devices because $\delta$ can set a minimum thickness of the conductive patch where the charge carriers can flow. $\delta$ indeed varies as a function of $\sigma$ according to the well-known relation:
\begin{equation}
    \delta = \sqrt{\frac{2\rho}{\omega\mu}} = \sqrt{\frac{1}{\pi\sigma\mu f}},
    \label{Eq:SkinDepth}
\end{equation}
\noindent where $\rho$ is resistivity, $\omega$ is an angular frequency, and $\mu$ is the permeability of the conductive film. 
The other criterion is the antenna quality factor that represents the losses of the antenna from various sources: due to the radiation loss ($Q_{\text{Rad}}$), due to the conduction loss ($Q_{\text{C}}$), due to the dielectric loss ($Q_{\text{D}}$), and due to the surface wave loss ($Q_{\text{SW}}$). The total quality factor ($Q_{\text{tot}}$) is described as \cite{Carver_1981,Balanis_2016}:
\begin{equation}
    \frac{1}{Q_{\text{tot}}} = \frac{1}{Q_{\text{Rad}}} + \frac{1}{Q_{\text{C}}} + \frac{1}{Q_{\text{D}}} + \frac{1}{Q_{\text{SW}}}.
    \label{Eq:Q-factor}
\end{equation}
While the optimized $Q_{\text{Rad}}$ is a function of $\epsilon_{\text{r}}$ and $h$, $Q_{\text{C}}$ is directly related to the skin depth effect, both will be discussed in the results section. 
Note that the $Q_{\text{D}}$ solely relies on material parameters, and $Q_{\text{SW}}$ is negligible because the loss from the surface wave is very small in microstrip antennas.  

To our best knowledge, extensive numerical simulations to assess various factors of conductive films and substrates have been elusive, and systematic investigation with quantitative analysis is absolutely required for microstrip antenna based on novel conductive nanomaterials, i.e. CNTs. Here, we perform systematic studies with a number of crucial material constant and dimension factors relevant for fabricating a thin, flexible microstrip antenna using COMSOL Multiphysics\textsuperscript{\textregistered}. First, we start by varying both $\sigma$ and thickness ($t$) of the patch conductive material. Next, we attempt to simulate realistic CNT conductive films in which the film $\sigma$ is spatially non-uniform due to several origins such as random alignment, non-uniform film thickness, and mixing with lower conductive SWCNTs. In an effort to mimic inhomogeneous CNT films in practice, we assess 7 artificial patterns, some of which can generate higher harmonic responses. Finally, we examine the dependence of the substrate thickness ($h$) and different $\epsilon_{\text{r}}$ as a practical guideline to select appropriate substrates for CNT-film strip antennas. Section II describes our numerical simulation setup and parameters, followed by the results and discussions in Sec. III. Based on our numerical studies, we propose a design guideline for microwave CNT film microstrip antennas with flexible and durable features in Sec. IV.

\section{Numerical Simulations}
    \subsection{Design Parameters of microstrip Antennas}
        
\begin{figure}[!hbtp]
        \centering
        \includegraphics[width=1\linewidth]{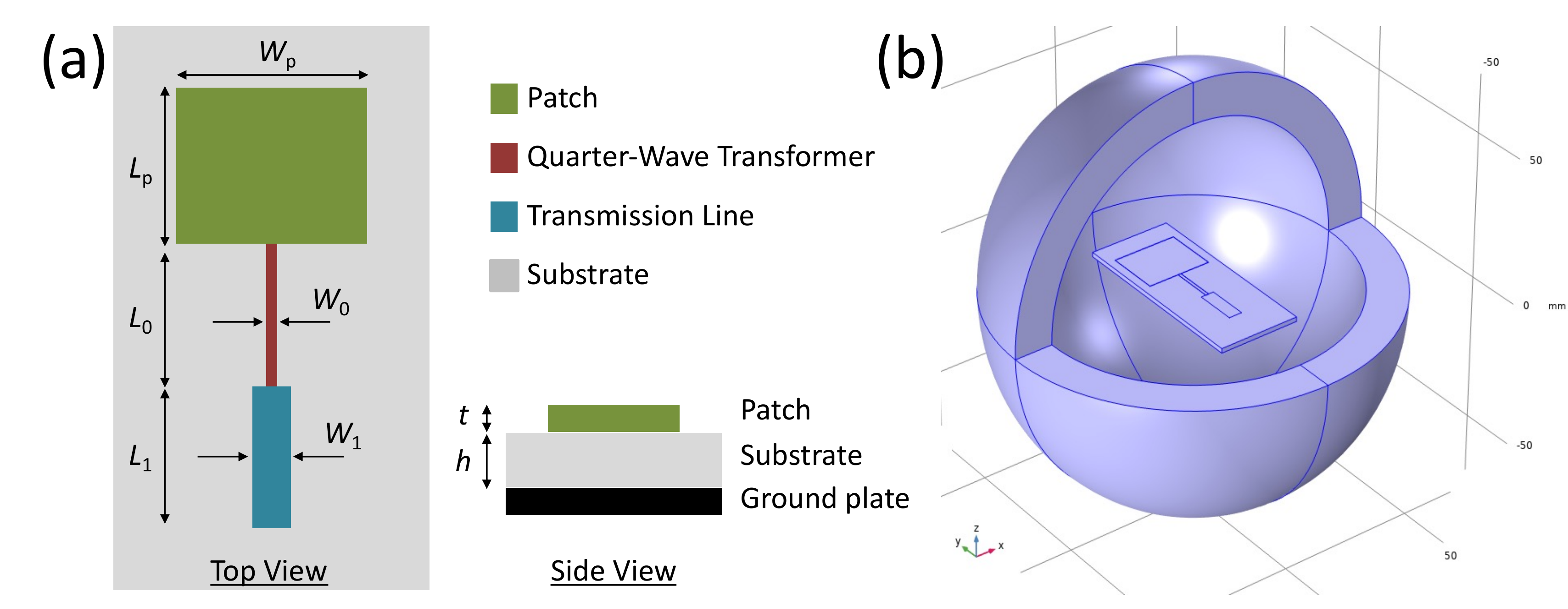}
        \caption{(a) (Left) A simple diagram of the basic antenna design used for simulations. Each part consists of width ($W$) and length ($L$) with subscripts: `p' for the microstrip antenna (green), `0' for the quarter-wave transformer (brown), and `1' for the transmission line (sky blue). (Right) The substrate (grey) and the conductive patch (green) on top of the ground plate (black) in the side view. (b) An isometric view of the antenna model in the COMSOL Multiphysics\textsuperscript{\textregistered} ~with one-quarter of the perfectly matched spherical boundary removed to make the microstrip antenna on the substrate visible inside.}
        \label{fig1:antenna_model}
    \end{figure}

At a given resonance frequency, the initial design parameters of a microstrip antenna are geometric dimensions of the simple planar structure and quantities of material parameters. A main microstrip antenna design (green rectangle in Fig.~\ref{fig1:antenna_model}(a)) has a physical width $W_{\text{p}}$ and a length $L_{\text{p}}$. $h$ is the thickness of the underlying substrate on which the microstrip antenna resides. A transmission-line model \cite{Balanis_2016} often determines the values of $W_{\text{p}}, L_{\text{p}}$, and $h$ at the fixed antenna frequency. The input feed of our antenna is broken into two rectangular sections: one serves as a quarter-wave transformer (brown area in Fig.~\ref{fig1:antenna_model}(a)), which is parameterized by a width $W_{\text{0}}$ and a length $L_{\text{0}}$, whereas the other section acts as a transmission line defined by a width $W_{\text{1}}$ and a length $L_{\text{1}}$ colored in blue of Fig.~\ref{fig1:antenna_model}(a). We confirm that these altogether satisfy the necessary 50 $\Omega$ impedance matching. Fig. ~\ref{fig1:antenna_model}(b) illustrates our simulation setup in the COMSOL Multiphysics\textsuperscript{\textregistered}, where we place the device structure at the center of a sphere, which sets the computation boundary for the far-field radiation patterns. 

\subsection{Simulation parameters of the conductor and the substrate }

Our numerical simulations investigate the CNT film antenna performance by varying geometric and material parameters of the conductor and the substrate: the conductor thickness $t$, conductivity distribution patterns, and substrate thickness $h$ with different dielectric constant values. The most commonly used microstrip antenna designs often neglect to consider the thickness of the conductive material by assuming it is ``very thin," i.e. $t \ll \lambda_0$ where $\lambda_0$ is the free-space wavelength \cite{Balanis_2016}. To our best knowledge, we have not found any literature to discuss the minimum thickness of the conductor explicitly. Regarding the minimum thickness, we must take into account the skin depth effect, which imposes a condition that $t$ should be adequately chosen by the $\sigma$-frequency relation with the Eq. (\ref{Eq:SkinDepth}), especially when $\sigma$ is not as high as $\sigma$ of metals. For the conventional Cu microstrip antennas, $t$ does not need to be considered because of extremely high conductivity $\sigma_\text{Cu} = 5.96 \times 10^7$ S/m, which yields small $\delta$.  Suppose a 5 GHz Cu antenna. $\delta$ = $0.919~\mu$m is definitely much thinner than standard Cu film thickness of the antenna (e.g. RT/Duroid\textsuperscript{\textregistered}~5880, thickness = 9 - 70 $\mu$m). Table~\ref{tab1:SkinDepth} compares the $\sigma$-dependent $\delta$ at 5 GHz. For materials having a conductivity between $10^6$ - $10^7$ S/m, $t$ is no longer a key design factor because the conductive film has often enough thickness to accommodate the skin depth effect. However, when lower conductive materials are exploited to make microstrip antennas, the skin-depth effect should be carefully examined to define the conductive film thickness. In our simulations, non-zero $t$ acts as a varying parameter to explore the skin depth effect on the performance of the antenna in a wide range of conductivity values (6 $\times 10^1$ S/m - 6 $\times 10^7$ S/m). As the AC current flows through a conductor, its thickness is compared with $\delta$ set by Eq. (\ref{Eq:SkinDepth}). The $t$ values are changed from 5 $\mu$m to 2400 $\mu$m. 

\begin{figure*}[!h]
        \centering
        \includegraphics[width=1.0\linewidth]{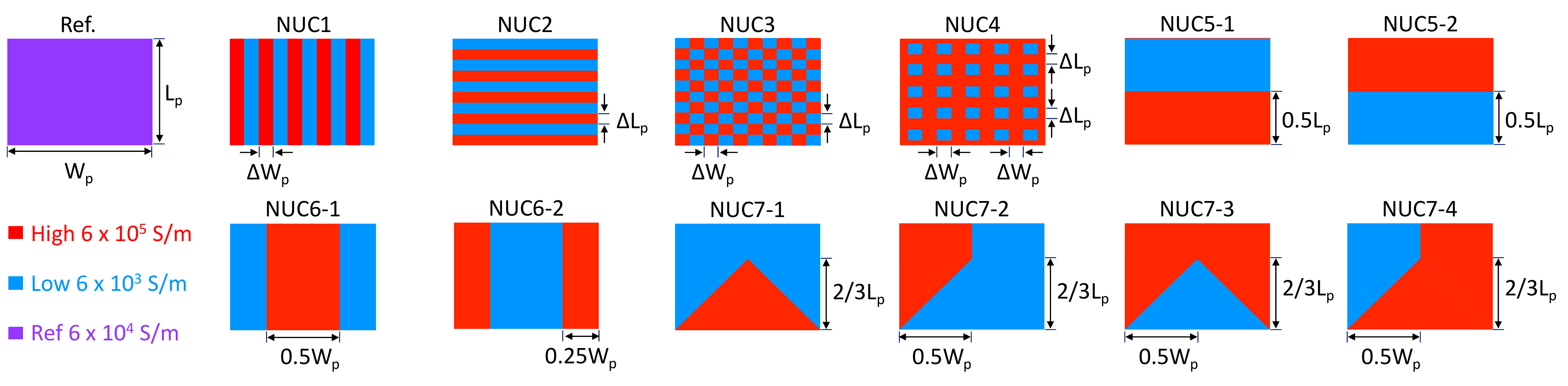}
        \caption{(a) The seven base designs for the non-uniform conductivity (NUC) antennas were simulated compared with the reference. NUC cases are designed with two different conductivity values, $6.0 \times 10^5$ S/m and $6.0 \times 10^3$ S/m, which are simplified possible non-uniform conductivity designs. $6.0 \times 10^4$ S/m can be considered as a reference which is a uniform structure. Non-uniform conductivity 1 (NUC 1) involves an alternating pattern of ten total alternating stripes along the x-axis, while NUC 2 consists of ten stripes along the y-axis. The third variant, NUC 3, combines these two into a checkerboard pattern, while NUC 4 contains isolated patches of low conductivity surrounded by high conductivity material. The latter three cases (NUC5 - NUC7) are designed to have large-scale non-uniformities inspired by the electric field spatial distributions at the fundamental, second, and third harmonic frequencies (details in Fig.~\ref{fig2:NUC5-7_ref}).}
        \label{fig2:NUCFamily}
    \end{figure*}

\begin{figure}[!hbtp]
        \centering
        \includegraphics[width=1.0\linewidth]{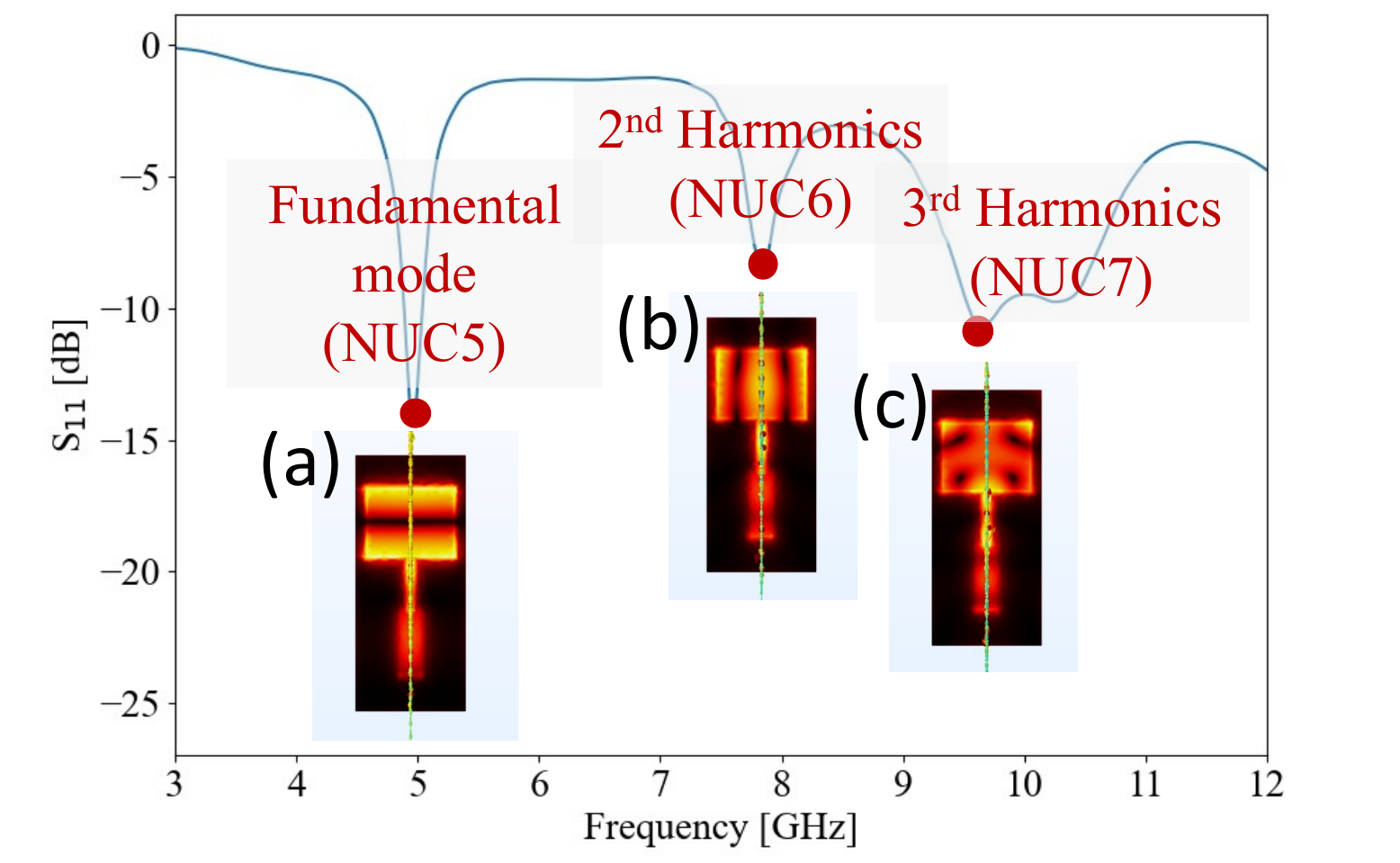}
        \caption{The reference patterns for NUC5 - NUC7 come from the resonance frequencies: three resonance frequencies are on the blue plot which is S$_{\text{11}}$. (a) the fundamental mode modeled to NUC5 (b) the second harmonics modeled to NUC6 (c) the third harmonics modeled to NUC7.}
        \label{fig2:NUC5-7_ref}
    \end{figure}

The second study is to explore the uniformity effect of the conductor film on the antenna performance with various patch patterns. Specifically, we make non-uniform conductor (NUC) patches mixed with high ($6.0 \times 10^5$ S/m) and low ($6.0 \times 10^3$ S/m) conductivity materials as an attempt to build a reasonable model for revealing possible non-uniformities in $\sigma$ that may arise in CNT nano-material films. The first four NUC cases (NUC1 - NUC4) shown in Fig.~\ref{fig2:NUCFamily}(a) are simplified to represent local, directional variations in horizontal, vertical, and both axes. The three remaining NUC cases (NUC5 - NUC7) mimic large-scale non-uniformities, whose $\sigma$-spatial configurations are inspired by the electric field distributions in the reference patch at the fundamental, second, and third harmonic frequencies (Fig.~\ref{fig2:NUC5-7_ref}).

Lastly, the variations of substrate thickness with three-different $\epsilon_{\text{r}}$ are simulated in order to study the substrate thickness effect. This study is especially crucial to design flexible and stretchable antennas. In this case, we fix $\sigma$ and $t$ from the first two studies. 

To quantify the antenna performance, the quarter-wave transformer model is considered to evaluate S$_{\text{11}}$, $\eta$, resonance frequency ($f_0$), and full-width half maximum (FWHM).  Especially, $\eta$ at the fundamental resonance frequency is calculated by integrating the radiated power over the far-field boundary in Fig.~\ref{fig1:antenna_model}(b).

\section{Results \& Discussion}

 \subsection{Reference Design Examination}
 
\begin{figure}[!hbtp]
            \centering
            \includegraphics[width=1.0\linewidth]{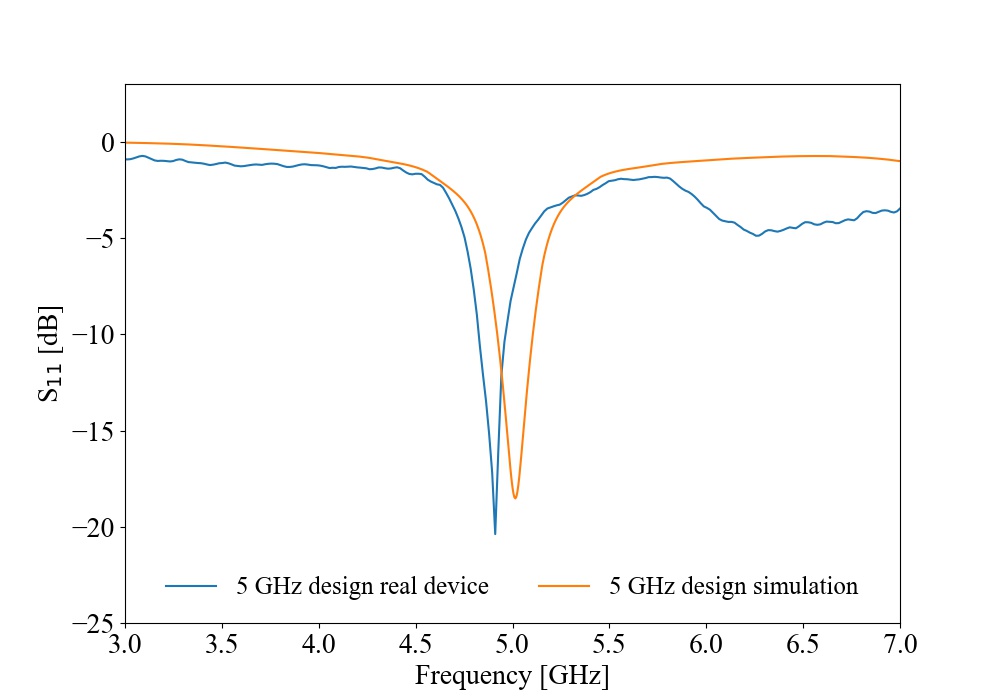}
            \caption{Comparing simulation and real device measurement result with reference 5 GHz design parameters. About 2.2\% for resonance frequency and 10 \% for S$_{11}$ differences are observed.}
            \label{fig3:Sim&real}
        \end{figure}

The reference simulation design targets a 5 GHz resonance frequency ($f_0$) with an RT/Duroid\textsuperscript{\textregistered}~5880 substrate ($\epsilon_r = 2.20$, $h$ = 1.575 $\text{mm}$), and the design parameters are specified in Table~\ref{tab:Sim&real_parameters}. A real antenna with the same design parameters in this reference model is fabricated, and the S$_{\text{11}}$ parameters are compared in Fig.~\ref{fig3:Sim&real}. Both results from the simulation and the real prototype device are plotted in orange and blue, respectively. The difference in $f_0$ between the simulated model and the real device is about 2.2 \% (simulation: 5.02 GHz, real device: 4.91 GHz), and the S$_{11}$ shows about 10 \% difference (simulation: -18.53 dB, real device: -20.38 dB). The measurement data from the real device are in excellent agreement with the simulation prediction, which provides us with confidence for further simulations to explore the aforementioned effects that are not easily controllable in real physical devices. 

        \begingroup
            \renewcommand{\arraystretch}{1.4}  
            \begin{table}[!hbtp]
                \centering
                \begin{tabular}{lcccccc}
                    \toprule 
                        & Simulation & Real Device \\
                    \midrule
                        resonance frequency [GHz]  & 5.02 & 4.91\\
                        S$_{11}$ [dB]  & -18.53  & -20.38 \\
                    \midrule
                        $L_p$ [mm] & 18.2 & 18.2 \\
                        $W_p$ [mm] & 23.7 & 23.7 \\
                        $L_0$ [mm] & 1.15 & 1.15\\
                        $W_0$ [mm] & 15 & 15 \\
                        $L_1$ [mm] & 17.5 & 17.5 \\
                        $W_1$ [mm] & 4.91 & 4.91 \\
                        $t$ [mm] & 0.07 & 0.07 \\
                    \midrule
                        $h$ [mm] & 1.575 & 1.575 \\
                        $\epsilon_r$  & 2.2 & 2.2 \\
                    \midrule
                        Cu conductivity [S/m] & 6x$10^7$ & $\sim$6x$10^7$ \\
                        Cu thickness [mm]  & 0.07 & $\sim$0.07 \\
                    \bottomrule  
                \end{tabular}
                \caption{Summary of performance data and antenna design parameters for simulation and real device in Fig.~\ref{fig3:Sim&real}. }
                \label{tab:Sim&real_parameters}
            \end{table}
        \endgroup

\subsection{Conductivity and Skin Depth}
  
\begin{figure}[!hbtp] 
            \centering
            \includegraphics[width=1\linewidth]{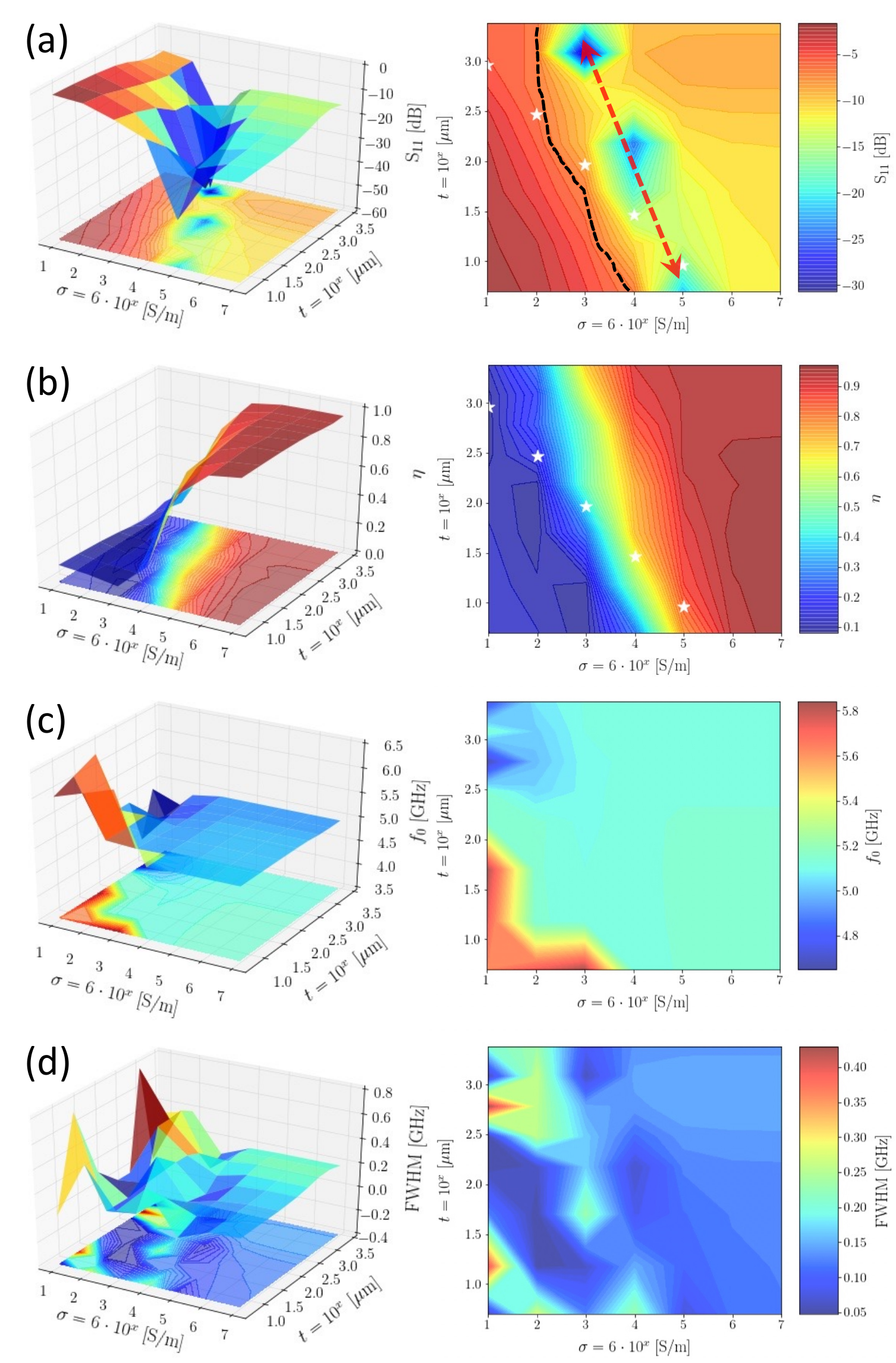}
            \caption{3D (left) and 2D (right) representation of results from the skin depth effect simulations as a function of $t$ conductor thickness and $\sigma$ conductivity in terms of (a) S$_{11}$ (the black dotted line is -10 dB contour line, the star marks are skin depth of each condition, and the red dotted arrow is a guideline for the lowest S$_{11}$ points), (b) radiation efficiency ($\eta$), (c) resonance frequency ($f_0$), and (d) full-width half-maximum (FWHM).}%
            \label{fig4:SkinDepth}%
    \end{figure}

The electric $\sigma$ of a single individual metallic CNT is known to be $\sim$ $10^6$ - $10^7$ S/m  almost as good as $\sigma_\text{Cu}$ = 5.96 $\times$ $10^7$ S/m or other metals. On the other hand, a random network of CNTs forming a film has a conductivity close to $10^2$ - $10^5$ S/m \cite{Wang_2018}, whose degradation is attributed to several material factors: the presence of lower $\sigma$ semiconductive CNTs, random orientations of nanotubes in films, and the abundance of interconnections between CNTs \cite{Li_2007,Zhou_2016}. Because of this lower conductivity, the skin depth effect should be taken into account for determining $t$ in CNT film microstrip antennas.  Table II calculates $\delta$ from Eq. (\ref{Eq:SkinDepth}) at 5 GHz for 7 different $\sigma$ values from 6 $\times$ $10^1$ to 6 $\times$ $10^7$ S/m, which covers the $\sigma$ ranges of both CNTs and Cu. For simulations, we set 8 different $t$ values, $t \in \{5, 15, 50, 150, 300, 600, 1200\}$ $\mu$m to evaluate four antenna metrics : S$_{11}$,  $\eta$, $f_0$, and FWHM in Fig.~\ref{fig4:SkinDepth}.
        
        \begingroup
            \renewcommand{\arraystretch}{2}  
            \begin{table}[!hbtp]
                \centering
                \begin{tabular}{cccccccc}
                    \toprule
                        & \multicolumn{7}{c}{Conductivity 6 $\times$ $10^x$ S/m} \\
                         $x$ & 1    & 2    & 3     & 4     & 5     & 6    & 7 \\
                    \midrule
                       $\delta$ ($\mu$m) & 919  & 291  & 91.9  & 29.1  & 9.19  & 2.91 & 0.919 \\
                    \bottomrule
                \end{tabular}
                \caption{Skin depth of a 5 GHz antenna for 7 conductivities from Eq.~(\ref{Eq:SkinDepth}).}
                \label{tab1:SkinDepth}
            \end{table}
        \endgroup
        
$\sigma$ also affects the antenna quality factor through the conductor contribution factor ($Q_{\text{c}} = h\sqrt{\pi \mu\sigma f} = h/\delta$ \cite{Carver_1981, Balanis_2016}), which can be readily captured in the S$_{\text{11}}$ and $\eta$ trends along with $t$ (Fig.~\ref{fig4:SkinDepth}). S$_{11}$ resonances are lower than an acceptable level -10 dB for $\sigma \geq 6 \times 10^4$ S/m with all $t$ values in Fig.~\ref{fig4:SkinDepth}(a). For the lower conductivity regions $\sigma \leq 6 \times 10^3$ S/m, $t$ must be increased to obtain S$_{11} < -10 $ dB, which can be explained by the previously mentioned 
skin depth effect (white star marks indicate skin depth $\delta$ at given $\sigma$ on Fig.~\ref{fig4:SkinDepth}(a) and (b)). In general, as $\sigma$ becomes higher and $t$ thinner, lower S$_{11}$ appears except two unique noticeable valleys indicated by the red dotted arrow.  Presumably, those points may be either due to high radiative efficiency with optimized $\sigma$ and $t$ at $f_0$ = 5 GHz or a possibility of high loss in the patch area or both together. A contour at  S$_{11}$ = -10 dB (black dotted line) sits in parallel to the valley line.  
When $t > 300$ $\mu$m at $\sigma = 6 \times 10^5$ S/m, S$_{11}$ becomes also shallower (Orange area in Fig.~\ref{fig4:SkinDepth}(a)), which is not optimal to operate. For the lower conductivity regions $\sigma \leq 6 \times 10^3$ S/m, the S$_{11}$ parameter rapidly becomes very shallow, thus $t$ must be increased to obtain S$_{11} < -10 $ dB in Fig.~\ref{fig4:SkinDepth}(a), which is understood by the aforementioned skin depth effect. The S$_{11}$ parameter approaches 0 dB as $\sigma$ nears 6 $\times$ 10 S/m for all $t$ values. 

Following the S$_{11}$ behavior, we observe a similar monotonic trend of $\eta$ in the $\sigma$ and $t$ domains (Fig.~\ref{fig4:SkinDepth}(b)). Given a fixed $t$, $\eta$ increases as $\sigma$ goes up. If $t$ gets thicker, the higher $\eta$ region appears. This $\eta$ trend somewhat overlaps with the S$_{11}$ behavior except for two S$_{11}$ valley regions. More efficient radiation expects less reflection, yielding lower S$_{11}$. However, the S$_{11}$ parameter is not necessarily lowest at the points of high $\sigma$ and high $\eta$. When $\sigma > 6 \times 10^5$ S/m, most of the power is radiated  over 90 \% regardless of the $t$ values. On the other hand, the loss mechanism is manifested in the lowest S$_{11}$ points. We suspect that the differences between S$_{11}$ and $\eta$ come from the loss mechanism in the patch region at the specific $\sigma$ and given $t$ around 5 GHz.
Another insightful observation in Fig.~\ref{fig4:SkinDepth}(a) is that the valley line (red-dotted arrow) in the $\sigma$ and $t$ phase spaces crosses the skin depth line near ($\sigma = 6\times 10^5$ \text{S/m}, t = 10 um), where $\eta$ is above 80 \%. This can be interpreted as the skin depth effect is irrelevant  in the higher $\sigma$ region based on the result of the $\eta$. In fact, the efficiency in 14 GHz devices by E. Amram Bengio \textit{et al.} becomes over 80 \% of $\eta$ regardless of $t$ similar to the $\eta$ of Cu antennas. On the other hand, $\eta$ of the lower $\sigma$ CNT antennas at 5 GHz and 10 GHz is much lower than the Cu $\eta$ at smaller $t$, which can by explained by skin depth effect \cite{Bengio_2019}. The $t$-independence in the $\eta$ in the actual CNT antennas made by Bengio \textit{et al.} is beyond our scopre here, requiring further studies. Hence, in order to fabricate an antenna made of CNT films or other nanomaterials for higher than 90 \% radiation efficiency, our overall results suggest that:
        
        \begin{itemize}
            \item $\sigma$ = 6 $\times$ $10^4$ S/m with $t$ $>$ 1,200 $\mu$m
            \item $\sigma$ = 6 $\times$ $10^5$ S/m with $t$ $>$ 50 $\mu$m
            \item $\sigma$ $>$ 6 $\times$ $10^6$ S/m for all $t >$ 5 $\mu$m.
        \end{itemize}  
       
Our simulations show that the variations of both $\sigma$ and $t$ have a minor impact on the $f_0$ of the antenna as seen in Fig. \ref{fig4:SkinDepth}(c)). They barely shift it from the original design value of 5 GHz within 1.1 \% error with the average $f_0$ and standard deviation, 4.97$\pm$ 0.06 GHz in all phase spaces excluding the low $\sigma$ and thick $t$ region (red left bottom corner in Fig. \ref{fig4:SkinDepth}(c)). Fig.~\ref{fig4:SkinDepth}(d) plots the FWHM results, which are relevant for designing a narrow-bandwidth patch antenna. Most of the configurations explored have fairly narrow bandwidth with the FWHM under -10 dB to be 2 \% of the $f_0$, which is consistent with the typical 2-5 \% FWHM of the resonance frequency in practical antenna designs~\cite{Balanis_2016}. We conclude that both $\sigma$ and $t$ do not affect the FWHM noticeably and no obvious trends of $\sigma$ and$\backslash$or $t$ on FWHM are observed.

\subsection{Non-uniform conductivity (NUC)}   
    
    \begin{figure}[!hbtp]
            \centering
            \includegraphics[width=1\linewidth]{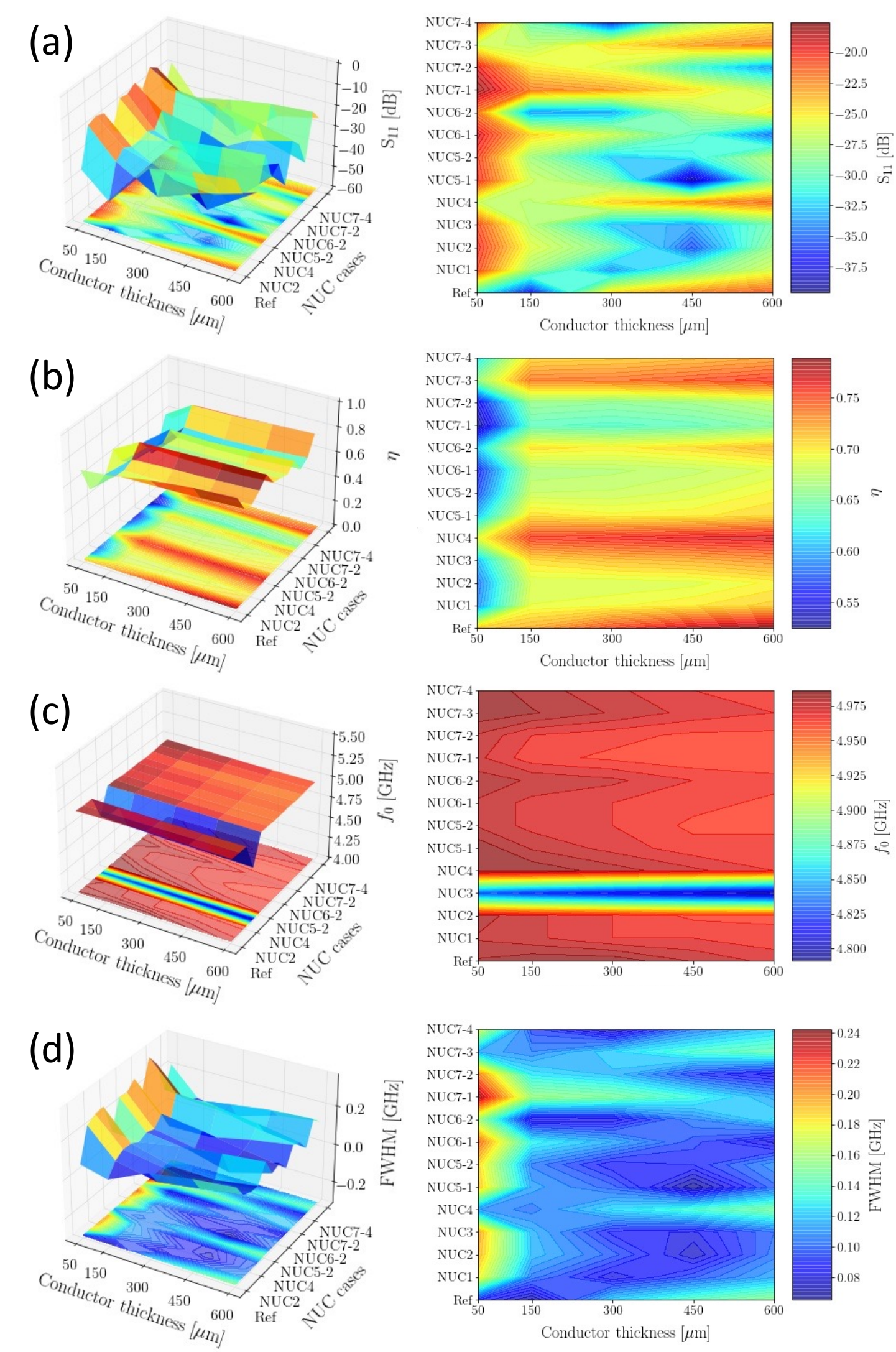}
            \caption{3D (left) and 2D (right) plots of non-uniform conductivity (NUC) simulations as a function of $t$ and 7 different NUC patterns with their variations with respect to the reference for (a) S$_{11}$, (b) radiation efficiency ($\eta$), (c) resonance frequency ($f_0$), and (d) full-width half-maximum (FWHM).
           }%
            \label{fig5:NUC}%
    \end{figure}    
        
Fig.~\ref{fig5:NUC} displays the simulation results of S$_{11}$, $\eta$, $f_0$, and FWHM for the NUC designs defined in Fig.~\ref{fig2:NUCFamily}. A reference is the uniform patch with 6 $\times$ $10^4$ S/m, whereas the NUC cases are patterned with the combination of 6 $\times$ $10^3$ S/m and 6 $\times$ $10^5$ S/m. First, the S$_{11}$ of NUC4, NUC6-2, and NUC7-3 tend to behave similar to that of the reference at $\sigma = 6 \times 10^4$ S/m, showing the degraded S$_{11}$ for thicker $t$ in Fig.~\ref{fig5:NUC}(a).
The common thing in three patterns is that the high $\sigma$ regions (6 $\times$ $10^5$ S/m) are situated at both left and right vertical edges in the patch, where the electric field radiates. This is an interesting result that the parallel components of the electric field propagate from the input port to the patch as a dominant role in both sides of the patch.

We can also identify the effect of the pattern directions between NUC1 (vertical) and NUC2 (horizontal), whose effective $\sigma$ is the same as 6 $\times 10^4$ S/m. Both NUC1 and NUC2 require thicker $t$ for reducing the reflection loss S$_{11}$ in order to compensate the lower $\sigma$ with respect to the reference. In addition, the horizontal NUC2 needs a much thicker $t$ than the NUC1 due to the similar directional reason of the above NUC4, NUC6-2, and NUC7-3. One 
distinctive observation is that in the checkerboard pattern of NUC3 with the 50:50 uniform combination of two different $\sigma$ values,  the S$_{11}$ follows the low effective $\sigma$ case, where the thicker $t$ helps to improve the S$_{11}$ monotonically. The last contributing factor in this study is the area fraction of the low $\sigma$ region in the patch. As expected, when the total area of the low $\sigma$ part is larger like in NUC7-1 and NUC7-2, the reflection loss can be compensated by making $t$ thicker.

The $\eta$ values of NUC4 and NUC7-3 are above 75 \% similar to the $\eta$ of the reference for all $t$ values, and  NUC3 has quite high $\eta$ of 75 \%. For the rest of the patterns in NUC1, NUC2, NUC5-1, NUC5-2, NUC7-1, and NUC7-2, $\eta$ can be improved by around 30 \% with $\sim$10 times thicker $t$. Therefore, given a NUC pattern, we select an appropriate $t$ by optimizing both quantities, S$_{11}$ and $\eta$. Higher $\eta$ is mainly followed by the large portions of high $\sigma$ in the patch as well as the pattern direction (Fig. \ref{fig5:NUC}(b)). It is already well-known that the $\sigma$ portion in the patch is readily expected to have an influence on the S$_{\text{11}}$ and $\eta$. According to our systematic simulations, surprisingly, the direction of the conductive films within the patch, furthermore the $\sigma$ value in both vertical edges are crucial as a controlling factor of the patch antenna properties. 

The pattern variations do not affect $f_0$ and FWHM in most cases shown in Fig.~\ref{fig5:NUC} (c) and (d), although the checkerboard NUC3 pattern antenna has $f_0$ = 4.8 GHz within 4 \% error from the original value.  
Most cases exhibit narrow FWHM, 0.2 \% of $f_0$ whose overall $Q$ = 25, a reasonable value. The $\eta$ values of NUC4 and NUC7-3 are above 75 \% similar to the $\eta$ of the reference for all $t$ values.

\subsection{Substrate thickness and dielectric constant variation}

\begin{figure}[!hbtp]
            \centering
            \includegraphics[width=1\linewidth]{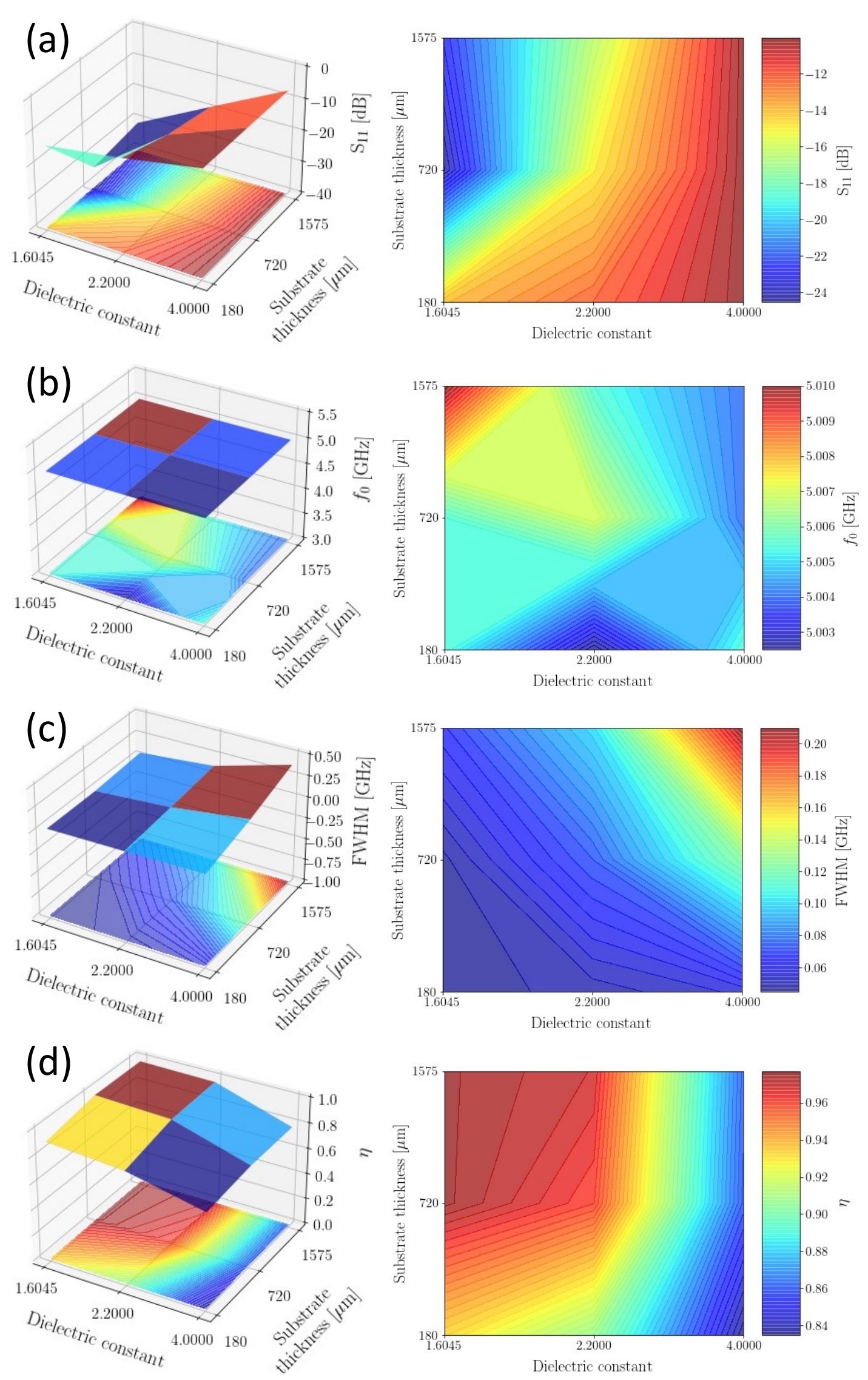}
            \caption{3D (left) and 2D (right) plots of substrate thickness variation simulations as a function of $\epsilon_{\text{r}}$ at $\sigma =$ 6 $\times$ $10^7$ S/m for (a) S$_{11}$, (b) radiation efficiency ($\eta$), (c) resonance frequency ($f_0$), and (d) full-width half-maximum (FWHM).} 
            \label{fig6:Substrate}%
        \end{figure}  

The simulations of the $h$ variation are performed with three representative materials of different $\epsilon_{\text{r}}$: 1.6045 for a sheet of paper or fabric, 2.20 for a PTFE plate, and 4 for a FR4 plate. Here we have a conductive film of $\sigma$ = 6 $\times$ $10^7$ S/m in the simulations. The $h$ values are set between 180 um and 1575 um to cover the thin substrate like paper and the thick PTFE case. From the $h$-dependence simulations displayed in Fig.~\ref{fig6:Substrate}, the minimum S$_{11}$ (Fig.~\ref{fig6:Substrate}(a)) is around -24 dB and $\eta$ (Fig.~\ref{fig6:Substrate}(b)) is larger than 84 \% in all values of $h$ for three substrates. The behavior of S$_{11}$ matches very well with the $\eta$ trend, where the lower S$_{11}$ and the higher $\eta$ one occur in the $h - \epsilon_r$ domains. In this case, it is good to confirm that less reflection loss means more radiation efficiency so that we can engineer $h$ and $t$ values for specific S$_{11}$ and $\eta$. Given a fixed $h$, smaller $\epsilon_r$ substrates win over larger $\epsilon_r$ counterparts, and thicker substrates further support lower S$_{11}$ and higher $\eta$ for smaller $\epsilon_r$ ones , which is well-matched with $Q_{\text{Rad}}$ ($Q_{\text{Rad}} \propto \frac{\epsilon_{\text{r}}}{h}$). Again $f_0$ is hardly affected by both $h$ and $t$ within 0.2 \% (Fig.~\ref{fig6:Substrate}(c)), and the extremely narrow FWHM of 0.1 \% of $f_0$ appears in sub-millimeter substrates (Fig.~\ref{fig6:Substrate}(d)). Thus, microstrip antennas on thin substrates would offer sharp, selective resonance responses, which are good for low noise and reliable channel links such as military or security required communications.

\section{Design guidelines for the CNT microstrip antenna}
    
Our systematic numerical studies give several important lessons that can be applied to design a flexible and durable antenna using a thin conductive film and a thin substrate based on CNT films whose conductivity is not as good as that of Cu. First, we have seen that the $\sigma$ of the film plays a significant role in the overall frequency response of the antenna, in the form of both resistive losses and losses due to the skin depth effect when it exceeds the thickness of the material. Above a certain value of $\sigma$, however, the antenna performance does not change significantly by $t$. According to our simulations, we conclude that for a radiation efficiency of about 90\%,  one should aim for a minimum $\sigma$ threshold of about  $6.0 \times 10^4$ S/m at $t > 1200~\mu$m. The film thickness can be reduced to be $t > 50~\mu$m for a conductive film of $\sigma = 6.0 \times 10^5$ S/m. If $\sigma$ is bigger than $6.0 \times 10^6$ S/m, any value of $t$ works.

The $\sigma$ uniformity in the conductive film plays an essential role to improve the $\eta$ from the simulations of the skin-depth effect. Considering the mixture of two different $\sigma$ materials, the larger areas of the high $\sigma$ material in a given antenna patch produces higher $\eta$. In particular, higher $\eta$ tends to appear for the patch designs where high $\sigma$ materials cover both vertical edges of the patch. One more observation is that small checker pattern does not significantly effect on the $\eta$, however, the pattern shifts about 4 \% on the $f_0$ to the lower frequency.
    
Lastly, we also examine how the thickness of the substrate affects the antenna performance which is directly relevant to designing a thin, flexible substrate. In general, thinner substrates are not preferable since they tend to suffer from a low radiation efficiency trend; however, we think that this may be compensated by adjusting the values of the dielectric constant. When the substrate is thinner, undesirable effects are seen on the S$_{11}$ and $\eta$ responses of the antenna, specifically in reducing the power absorbed by the antenna at the first resonance frequency.
    
Overall, we consider the factors discussed in Section III to reach design guidelines of flexible microstrip antennas using CNT films in the microwave frequency domain: A suggested target $\sigma$ value of the CNT conductive film should be at least $\sigma$ of $6.0 \times 10^5$ S/m with a minimum thickness of 50 $\mu$m. Making uniform $\sigma$ CNT films is a key factor and this can be achieved by aligning purified CNTs in films, that can be placed on a paper or a fabric as such low-dielectric flexible substrates.

\section{Conclusion}

We report our methodological analysis of several crucial design factors on the flexible microstrip antenna performance via numerical simulations in the COMSOL Multiphysics\textsuperscript{\textregistered}. Specifically, both $\sigma$ and $\delta$ of the conductive materials directly influence on the S$_{11}$ and the radiation efficiency of the antenna. Furthermore, various non-uniform spatial distributions of $\sigma$ are modeled together with the substrate parameters of the thickness and dielectric constant. Our numerical study allows us to design promising microstrip antennas based on CNT films for flexible, lightweight, and high-frequency microwave applications.

\section*{Acknowledgment}

This research was undertaken thanks in part to funding from the Ontario Research Fund-Research Excellence (ORF-RE) and the Canada First Research Excellence Fund - Transformative Quantum Technologies (CFREF-TQT). We acknowledge the support of Industry Canada, the Ontario Ministry of Research \& Innovation through Early Researcher Awards (RE09-068), and the CMC Microsystems for the CAD tools. The authors thank Safieddin Safavi-Naeini, John Long, Aidin Taeb for fruitful discussions and Steve Weiss for supporting computational resources.

\bibliographystyle{ieeetr}
\bibliography{refs}
\end{document}